# Ultrafast Internal Conversion Dynamics Through the on-the-fly Simulation of Transient Absorption Pump-Probe Spectra with Different Electronic Structure Methods

Chao Xu[1], Kunni Lin[1], Deping Hu[2], Feng Long Gu[1], Maxim F. Gelin[3] and Zhenggang Lan[2,*]

[1] Key Laboratory of Theoretical Chemistry of Environment, Ministry of Education; School of Chemistry, South China Normal University, Guangzhou 510006, P. R. China.

[2] Guangdong Provincial Key Laboratory of Chemical Pollution and Environmental Safety and MOE Key Laboratory of Environmental Theoretical Chemistry, SCNU Environmental Research Institute, School of Environment, South China Normal University, Guangzhou 510006, P. R. China.

[3] School of Sciences, Hangzhou Dianzi University, Hangzhou 310018, P. R. China.

**ABSTRACT:** The ultrafast nonadiabatic internal conversion in azomethane is explored by the on-the-fly trajectory surface-hopping simulations of photoinduced dynamics and femtosecond transient absorption (TA) pump-probe (PP) spectra at three electronic-structure theory levels, OM2/MRCI, SA-CASSCF, and XMS-CASPT2. All these dynamics simulations predict ultrafast internal conversion. On the one hand, the OM2/MRCI and SA-CASSCF methods yield similar excited-state dynamics, while the XMS-CASPT2 method predicts a much slower population decay. On the other hand, the TA PP signals simulated at the SA-CASSCF and XMS-CASPT2 levels show the similar spectral features, particularly for the similar stimulated emission contributions, while the OM2/MRCI signals are quite different. This demonstrates that the nonadiabatic population dynamics and time-resolved stimulated emission signals may reflect different aspects of photoinduced processes. The combination of the dynamical and spectral simulations definitely provides more accurate and detailed information which sheds light on the microscopic mechanisms of photophysical and photochemical processes.

**TOC Graphic**

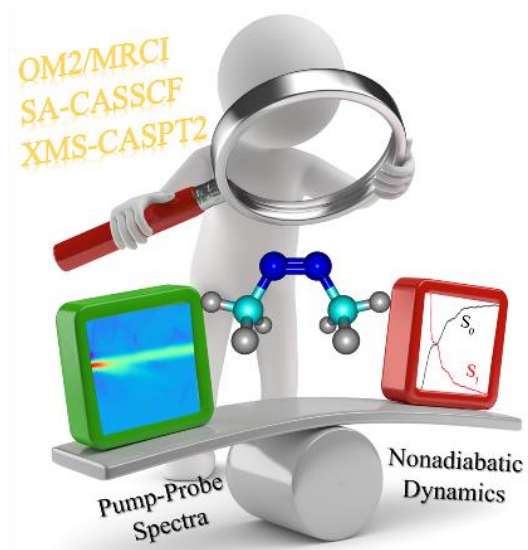

Light-driven non-adiabatic processes are essential in many photophysical, photochemical, and photobiological reactions.[1-6] A comprehensive understanding of these processes is fundamentally significant for uncovering the microscopic mechanisms of photoswitching[7,8], photosynthesis[9,10], and photostability[11,12].

Time-resolved optical spectroscopy is a powerful experimental technique to study the photoinduced non-adiabatic dynamics.[5,11-14] However, it is not straightforward to associate the directly measured spectroscopic observables with specific photochemical or photophysical processes without the pre-knowledge of excited-state dynamics therein. Therefore, the interpretation of spectroscopic signals relies heavily on the support from theoretical calculations and simulations.[14-20] The theoretical explanation of most time-resolved nonlinear spectroscopic signals is based on the third-order nonlinear optical response of the light-matter interaction.[21-23] In commonly-used transient absorption (TA) pump-probe (PP) experiments, the system is perturbed by a pump laser pulse at first, then the system's response after dynamical evolution is detected by another delayed probe pulse. The observed TA PP signal thus reflects the third-order response of the system on the laser fields of the pump and probe pulses, which gives the spectral fingerprint of the system dynamics governed by the time-dependent evolution of the density operators.[21-23]

In recent years, the trajectory-based on-the-fly *ab initio* non-adiabatic dynamics simulations are booming and are used extensively to investigate the ultrafast nonradiative electronic decay in polyatomic systems.[4,24-26] Most on-the-fly non-adiabatic simulations focus on recoding the dynamics evolution in the photoinduced processes. Correspondingly, their results give rich information related to time-dependent molecular geometries, energies, and electronic-state populations. Comparatively little attention has been paid to the simulations of experimentally measured time-resolved signals in the basis of on-the-fly non-adiabatic dynamics, although several methods and protocols have been developed for the simulation of PP-like signals. *Ab initio* simulations of time-resolved PP photoelectron spectra were carried out by using classical-trajectory guided Gaussian basis set method and semiclassical Wigner representation method.[3, 27] In these simulations, one evaluates, in fact, excited-state absorption (ESA) contribution to the TA PP signal, which yields projection of the vibronic wavepacket in the lower-lying electronic states to the cationic product state. The time resolved fluorescence spectrum, which has been simulated by employing the Einstein coefficient approach,[28-33] is proportional to the stimulated-emission (SE) contribution to the TA PP spectrum. On-the-fly simulations of TA PP spectra were performed by employing the cumulant/harmonic approximation for vibrational contributions,[34-36] single-Gaussian-wavepacket Ansatz[37] and real-time time-dependent density-functional theory methodology.[38-40] In aforementioned simulation methods a number of simplifying assumptions are unavoidable, since one must approximate essentially quantum dynamics of the molecular system driven by external fields by evolution along classical trajectories. Hence novel approaches and protocols, notably those allowing the on-the-fly simulation of TA PP signals with realistic pulse envelopes are still necessary. For instance, Gelin, Domcke, and co-workers recently proposed a practical approach to simulate the TA PP signals by combing the on-the-fly trajectory simulation and the doorway-window (DW) representation of nonlinear spectroscopy.[41]

While the development of the universally applicable and reliable tools for the simulation of TA PP signals of polyatomic chromophores and molecular aggregates is certainly challenging, the other theoretical issue is also significant in applications of such simulations to specific molecular systems. It is well known that nonadiabatic dynamics simulations with different electronic structure methods often deliver different or even contradictory results.[42, 43] However, benchmarking of time-resolved TA PP spectra simulated at various quantum-chemistry levels is still lacking. Herein, the aim of this work is to investigate the performance of three different electronic structure methods in the nonadiabatic dynamics simulation of TA PP spectra and in the monitoring of internal conversion to the electronic ground state. For this purpose, we performed on-the-fly *ab initio* non-adiabatic trajectory surface hopping (TSH) dynamic simulations[44] and evaluated TA PP signals of azomethane (AZM in Figure S1) at different electronic structure levels. The OM2/MRCI, SA-CASSCF, and XMS-CASPT2 methods were employed here because they can well describe the topology of $S_0/S_1$ potential energy surfaces. As AZM is the simplest azoalkane, the study of photodynamics processes in this molecule has attracted considerable attention[45-49]. The first low-lying $n$-$\pi$* excited state was found to play a crucial role in the photoinduced decay. Many experimental and theoretical studies have been conducted to investigate the photoinduced dynamics of AZM after photoexcitation. It was scrutinized by femtosecond-multiphoton spectroscopy,[50] nanosecond-coherent anti-stokes Raman spectroscopy[51], *ab initio* potential energies surface calculation, and dynamics simulation.[49, 52, 53]

In the present work, the TA PP signals were simulated with the DW representation of the nonlinear response.[21, 41, 54-58] This method, which is valid if the pump and probe pulses are temporally well separated, allows one to fully account for realistic pulse shapes in the time and frequency domain. The theoretical aspects of the method and interfacing of the DW representation and the TSH simulation procedure were discussed in Ref[41, 59]. Here we extend this theoretical framework to take into account the $S_1 \rightarrow S_0$ internal conversion, thereby allowing trajectories in the lowest excited electronic state $S_1$ jump to the ground electronic state $S_0$ (see Supporting Information for all theoretical and computational details). Then we study how the simulated photoinduced dynamics and TA PP spectra depend on the use of different levels of electronic-structure theories.

We begin with the population dynamics in Figure 1, which shows the time-dependent fractional occupations of the $S_0$ and $S_1$ electronic states of AZM, assuming that all trajectories started from the $S_1$ state. Very similar population evolutions are obtained at OM2/MRCI[60] and SA-CASSCF[61, 62] levels, giving the 50% $S_1$ decay around 91 fs and 105 fs, respectively. At 400 fs, the $S_1$ population becomes very low, and the $S_1 \rightarrow S_0$ internal conversion is essentially over. However, the TSH dynamics at the XMS-CASPT2[63] level yields a much longer time scale (249 fs) of the 50% $S_1$ population decay. At 400 fs, more than 20% of the population remains in the $S_1$ state.

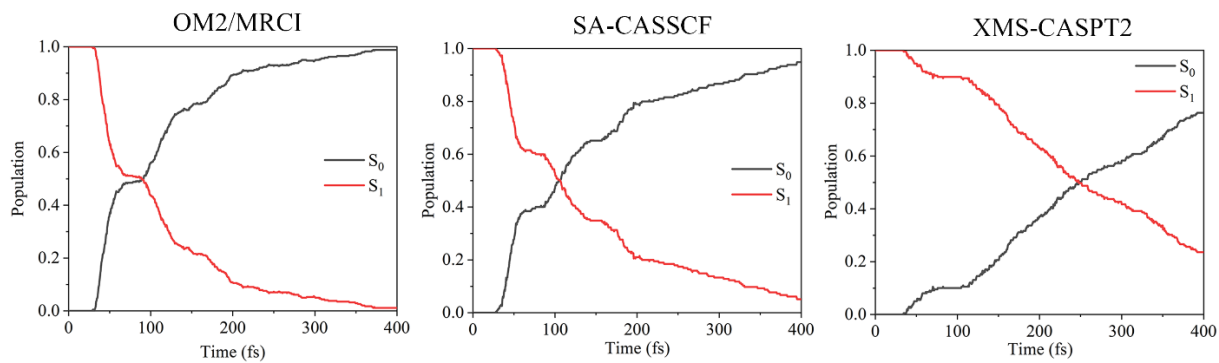

**Figure 1**. Time-dependent fractional occupations of the $S_0$ and $S_1$ electronic states of AZM in non-adiabatic dynamics staring from the $S_1$ state. 200 trajectories were used for obtaining the converged results.

Next, let us consider TA PP signals. It is appropriate to start from the qualitative picture of the formation of these signals in AZM. The total TA PP signal can be decomposed into three contributions, the ground-state bleach (GSB), SE, and ESA. In the DW formalism, these contributions are calculated as follows. At $t=0$, the pump pulse creates the Doorway-wavepackets (D-wavepackets) in the states $S_0$ and $S_1$. Then, classical trajectories are propagated in the states $S_0$ and $S_1$ up $t=\tau$ where $\tau$ is the time delay between the pump and probe pulses. At $t=\tau$, the probe pulse creates the Windows-wavepackets (W-wavepackets), which probe the nuclear motion of the $S_0$ (GSB by $S_0 \rightarrow S_1$), $S_1$ (SE by $S_1 \rightarrow S_0$, ESA by $S_1 \rightarrow$ higher-lying). Finally, the TA PP signal is obtained by averaging of the product of the D-wavepacket at $t=0$ and the W-wavepacket at $t=T$ over nuclear trajectories.[41] In the simulation of TA PP signals, the pump and probe pulses with Gaussian envelopes $E(t) = \exp\{-(t/\tau_P)^2\}$ were used. The pulse durations for both the pump and probe pulses are set to 5 fs, which yields the bandwidth of 0.44 eV (full width at half-maximum).

The study of the $S_1 \rightarrow S_0$ internal conversion in AZM requires the modification of the DW simulation protocol of Ref[41]. Within the TSH picture, a trajectory contributes to the SE and ESA signals only when it propagates in the excited state $S_1$. When the trajectory jumps back to the ground state $S_0$, its contribution to the SE and ESA signals vanishes while the contribution to the GSB signal emerges. Hence, the GSB signal consists of two components. The first (hereafter "cold") component is the conventional GSB signal which reflects the evolution of the hole wavepacket created and probed in $S_0$. The second (hereafter "hot") component describes the wavepackets which are launched by the pump pulse into $S_1$, undergo $S_1 \rightarrow S_0$ internal conversion and are interrogated by the probe pulse in $S_0$. As explained in Supporting Information, the two components of the GSB signal have opposite signs. Following the convention used in the present work, the cold GSB and the SE signals are positive, while the hot GSB and the ESA are negative.

The GSB, SE, and ESA contributions as well as the total integral TA PP signal within the first 300 fs are shown in Figure 2. To enhance the short-time features, the same signals are depicted on the time scale of 50 fs in Figure S2.

The GSB signals are plotted in the upper panels of Figure 2. As explained above, the GSB signals consist of two components. The cold (positive) component is almost stationary. It is located around $\omega_{pr}$ ~ 3.4 eV (OM2/MRCI), 3.7 eV (SA-CASSCF), and 3.4 eV (XMS-CASPT2). The hot (negative) high-amplitude component emerges around 50 fs (OM2/MRCI) and 100 fs (XMS-CASPT2) and oscillates around the cold component.

The SE contribution (middle panels) directly reflects the wavepacket evolution in the $S_1$ state. The SE spectra computed at the SA-CASSCF and XMS-CASPT2 levels are relatively similar, except for the different original emission positions. Let us take the SA-CASSCF signal as an example. The excited-state trajectories are initiated in the $S_1(n\pi^*)$ state, because the excitation energy of this state is assumed to be in resonance with the central frequency of the pump pulse. The signal moves to the lover frequencies, significantly weakens already at $\tau$ ~ 10 fs and almost vanishes at $\tau$ ~ 30 fs. However, OM2/MRCI provides a different evolution of the SE component. Within the first 20 fs the SE signal moves to the lower frequencies and its intensity becomes slightly weaker. However, within 25-40 fs, the SE intensity shows a recurrence and then decays almost completely on the time scale of around 150 fs. During this evolution, the SE signal shifts to the low-energy domain.

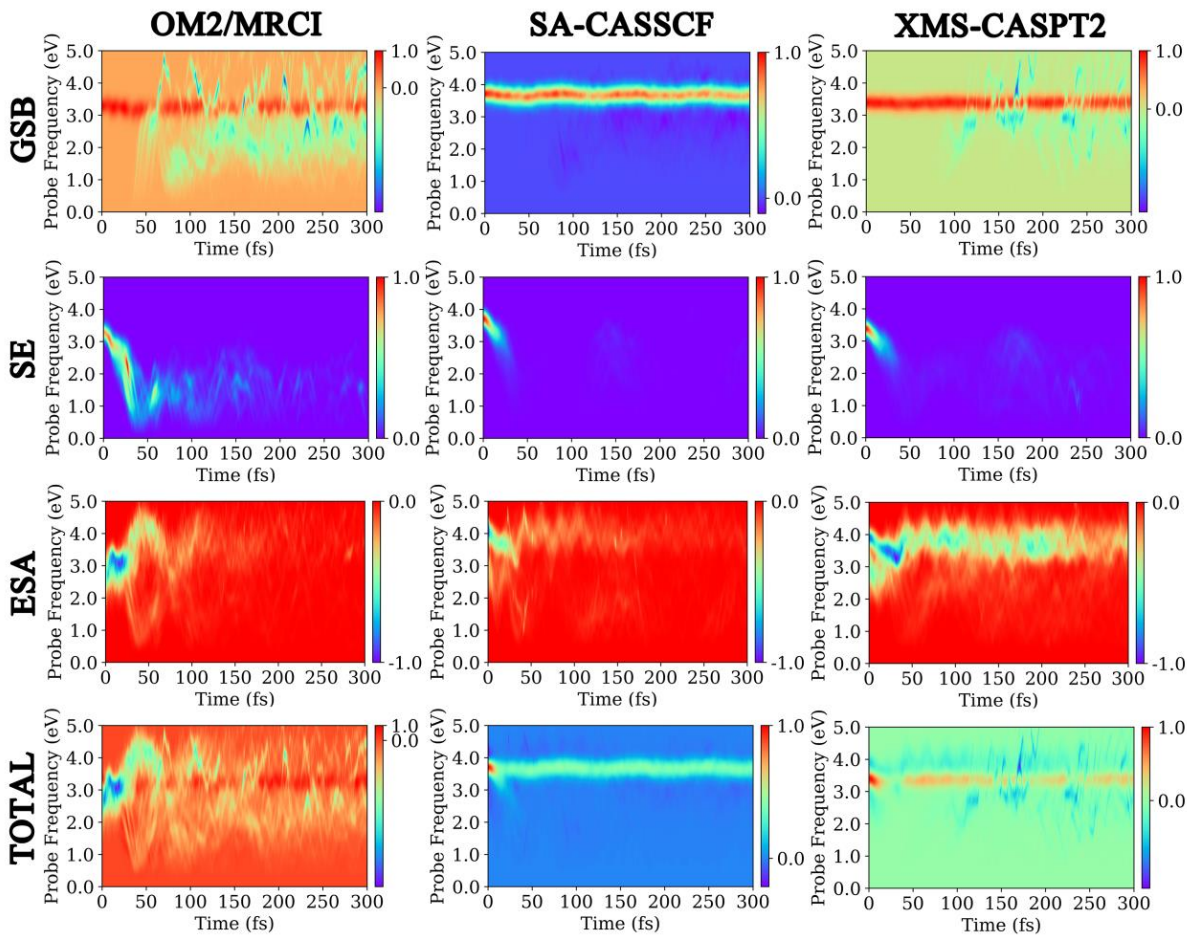

**Figure 2.** Normalized GSB, SE, and ESA contributions and total integral TA PP signal of AZM as a function of the pump-probe delay time T and the central frequency $\omega_{pr}$ of the probe pulse. Duration of both pump and probe pulses is 5 fs. The central frequency $\omega_{pu}$ of the pump pulse is tuned into resonance with the $S_1(n\pi^*)$ state (OM2/MRCI: 3.37 eV, SA-CASSCF: 3.70 eV, XMS-CASPT2: 3.40 eV). Left column: OM2/MRCI method, Middle column: SA-CASSCF method, Right column: XMS-CASPT2 method.

The ESA signals reflect the electronic transitions from the lower-lying excited states (in which excited-state wavepacket dynamics takes place) to the optically-allowed higher-lying excited states. It is not trivial to simulate ESA spectra, as the excitation from the current dynamically-relevant excited state to higher-lying excited states must be calculated. Hence the ESA contributions presented in Figures 2 and S2 can be regarded as semi-quantitative, since they are calculated by using snapshots taken from the trajectory propagation and performing the additional single-point calculations of four electronic states at individual electronic-structure level. The ESA signals given at the OM2/MRCI level are slightly red-shifted in the very beginning and then exhibit the significant blue shift by ~ 1.5 eV at $\tau > 50$ fs. This indicates that the optically-allowed electronic transition from the $S_1$ to higher-lying excited states become possible along the trajectory. When more trajectories jump back to the electronic ground state, the ESA signal almost disappears at $\tau > 200$ fs. At the SA-CASSCF and XMS-CASPT2 levels, different ESA signals are obtained, while both spectral shift and intensity decay are given within 50 fs. The ESA signals vanish in the longer time scale owing to the $S_1 \rightarrow S_0$ internal conversion.

At the XMS-CASPT2 level, the nonadiabatic population decay is slower than that at other two levels (see Fig. 1) and the corresponding ESA signal also vanishes on the longer time scale. The significant differences in the ESA signals are attributed to the fact that the high-lying excited states responsible for the ESA signals significantly depend on the level of electronic structure theories. It is well known that such dependence represents a great challenge in quantum chemistry.

The total integral signal is the sum of the GSB, SE, and ESA contributions, as depicted in Figures 2 and S2. The OM2/MRCI TA PP spectrum is initially dominated by the ESA signal, but later also exhibits clear hot and cold GSB contributions. The SA-CASSCF signal is dominated by the cold GSB contribution, and the XMS-CASPT2 signals exhibits pronounced GSB and ESA contributions. It is important that the SA-CASSCF and XMS-CASPT2 spectra in the low energy region are similar, and the early-time decaying (at ~ 20 fs) SE contributions may be extracted from the total TA PP signal, when we consider the frequency shift of the SE signal.

No matter which electronic structure methods are used, both on-the-fly TSH nonadiabatic dynamics and time-resolved TA PP signals indicate that the internal conversion in AZM takes place on the ultrafast time scale. However, the detailed comparison of the nonadiabatic population dynamics in Figure 1 and the TA PP signals in Figure 2 reveals a number of interesting questions. For example, OM2/MRCI and SA-CASSCF give the similar population dynamics, while the corresponding TA PP spectra differ substantially. At the same time, SA-CASSCF and XMS-CASPT2 predict noticeably different $S_0$ and $S_1$ population evolutions, but the corresponding TA PP signals, in particular the GSB and SE components, show many similarities. In addition, simulations at all three levels of electronic structure theory predict that the SE signals decay much faster than the $S_1$ populations. Usually, it is assumed that the time-resolved SE signal directly reflects the nonadiabatic population decay, while our results indicate that the two observables are not equivalent for AZM. Below we analyze all these intriguing problems in detail.

To understand the difference between the predictions of the three electronic structure methods, the reaction pathway from the ground-state minimum ($S_0$_min) to the conical intersection was built by employing the linear interpolated method, and the corresponding potential energy curves as functions of the dihedral angle C-N-N-C obtained by the three methods are presented in Figure 3(a). The three methods predict that the barrierless pathways on the $S_1$ potential energy curve connect the Franck-Condon (FC) point with the $S_0/S_1$ conical intersection. As a function of the torsional angle, the XMS-CASPT2 pathway is flatter compared to the OM2/MRCI and SA-CASSCF pathways. This clearly explains why the TSH dynamics at the XMS-CASPT2 level predicts the longer $S_1$ excited state lifetime. The $S_0\_min$ geometry and the excited state properties at $S_0\_min$ are given in Supporting Information.

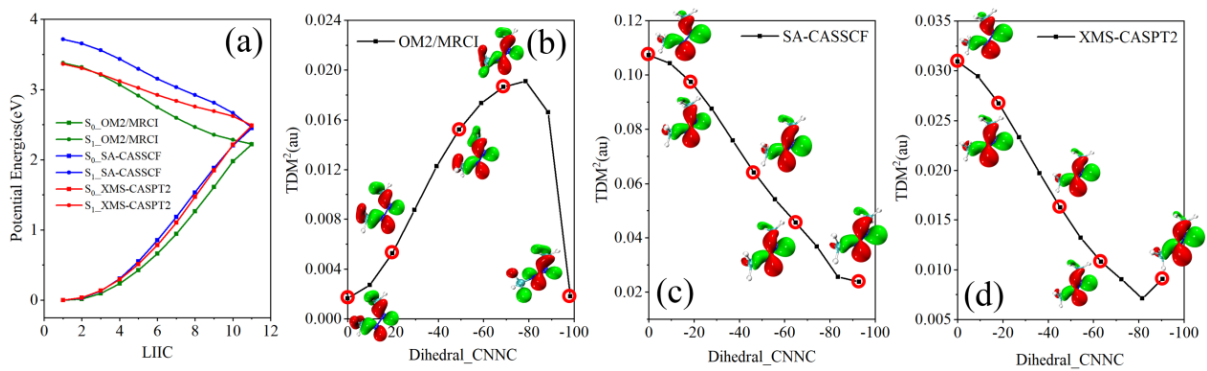

**Figure 3.** Potential energy curves (a), $S_0$-$S_1$ TDMs squared and the highest occupied molecular orbitals (contributing to the electronic transition in the $S_1$ state) along the linear interpolated pathway from $S_0\_min$ to the conical intersection, (b)-(d), calculated by the electronic-structure methods indicated in the legends.

The calculated TDMs between the $S_0$ and $S_1$ states and the highest occupied molecular orbitals along the linear-interpolated excited-state reaction pathway from the ground state minimum to the $S_0/S_1$ conical intersection are given in panels (b)-(d) of Figure 3. As discussed above, the linear interpolated potential energy curves calculated by the three methods are quite close to each other. However, the $S_0$-$S_1$ TDMs depend dramatically on the chosen levels of the electronic structure theories. For instance, both SA-CASSCF and XMS-CASPT2 methods predict that the TDM decreases monotonically with the C-N-N-C dihedral angle from the FC region to the $S_0/S_1$ conical intersection. Contrastingly, the different dependence of the TDM on the torsional angle was obtained at the OM2/MRCI level. The TDM at the $S_0$_min is small, then it increases, reaches its maximum at around 70°, further decreases with the torsional angle, and finally nearly vanishes at the conical intersection. This difference can be well explained by the electronic character of the highest occupied molecular orbital (HOMO) contributing to the electronic transition of the $S_1$ state. Essentially, this is a *n* orbital, while we observe the appearance of the small σ component in the HOMO at the OM2/MRCI level. Such *n*-σ mixing does not appear at the other two levels. Since the σ component displays much lower overlap with the π* orbital density, the TDM of the $S_1$ state at the OM2/MRCI becomes much smaller. The different $S_0 \rightarrow S_1$ TDM features and the non-Condon effects described clearly explain the remarkable differences in the SE and GSB signals obtained at the OM2/MRCI, SA-CASSCF, and XMS-CASPT2 levels.

Let us now discuss the SE component of the TA PP signal. In the early stage of the dynamics, the trajectory experiences the torsional motion of the CNNC dihedral angle on the $S_1$ potential energy surface, as shown in Figure S3. Both the SA-CASSCF and XMS-CASPT2 predict that the TDMs decrease dramatically due to the fast CNNC torsional motion. Along this torsional motion, the $S_0$-$S_1$ energy gap also becomes smaller. As the consequence, the SE signals first appear in the FC region at the beginning of the nonadiabatic dynamics and then move to the lower-energy domain. At the same time the SE intensity becomes weaker and quickly vanishes when the trajectory enters the regions in which the TDMs become rather small. We also noticed that the vanishing of the TDM values begins much earlier than the trajectories access the $S_0/S_1$ conical intersection. Therefore, the SE signals quench much earlier than the $S_1$ populations. When the trajectory is propagated at the OM2/MRCI level, the torsional motion increases the TDM and decreases the $S_0$-$S_1$ energy gap. Thus, the SE signal moves to the low-frequency domain in this early stage. Although the TDM increases, different trajectories start to spread over a broad area in the phase space, and these trajectories show different transition energies. As the result, the SE intensity does not increase significantly. When most trajectories move close to the geometries with larger TDMs at a CNNC dihedral angle of ~ 60 ° at about 25 fs, the recurrence of the SE appears in the low-energy domain. Next the trajectories move to the regions with small $S_0$-$S_1$ TDM at 30-40 fs and the SE signal starts vanishing. As distinct from the SA-CASSCF and XMS-CASPT2 results, the recurrence of the SE signal is also attributed to the non-vanishing $S_0$-$S_1$ TDM along the torsional motion at the OM2/MRCI level. Overall, the differences discussed here indicate that the OM2/MRCI method may not be accurate enough to describe the excited-state electronic wavefunction and TDM for the current system, even if the state energy may be reasonable.

The time-dependent state population decay is governed exclusively by the profiles of the excited-state pathways. The SE signal quenches both due to the vanishing of TDMs in certain areas of the nuclear phase space and the internal conversion to the ground state. This explains why the SE signals decay much faster than the nonadiabatic populations and demonstrates the nontrivial interconnection between the nonadiabatic dynamics and the TA PP signal evolution.

Next, let us analyze the GSB signal in details. The "cold" component in the GSB signal is relevant to the dynamics of the "hole" created by the D-wavepacket. The weak modulation here is caused by the combination of non-Condon effects in the Frank-Condon region (Figure S4), anharmonic effects, and finite pulse duration. The "hot" GSB signal is governed by the interplay between to the "hot" vibrational motion and the strong dependence of the $S_1$-$S_0$ TDM on nuclear coordinates (Figure 3). The $S_1$-$S_0$ TDM decreases dramatically with the torsional motion at the SA-CASSCF and XMS-CASPT2 levels. It should be noted that the current simulations are performed for isolated AZM. Hence the excessive energies drive the large-amplitude nuclear motion, and the spreading of the $S_1 \rightarrow S_0$ trajectories in the phase space makes the hot component of the GSB signal quite erratic. In the condensed phase, the excessive energies of the system are gradually dissipated by the environment, so that the system finally equilibrates and returns back to the ground state minimum. In this situation, the two GSB components should cancel each other and the total GSB signal should eventually disappear. This is a well-known feature in the TA PP experiments (the so-called return of transparency), but this vibrational cooling occurs on a much longer (from tens to hundreds of picoseconds) time scale. We also noticed that the SA-CASSCF and XMS-CASPT2 give different dependences of the $S_1$-$S_0$ TDM on the reactive coordinate, and the former level gives much faster decay, as shown in Figure 3. After the internal conversion, many trajectories still remain in the low TDM region. Since the W-function is proportional to the square of TDM, the hot GSB signal is very weak at the SA-CASSCF level in Fig. 2. As the contrast, the OM2/MRCI level shows that the $S_1$-$S_0$ TDM is very small at $S_0$ minimum, and thus the hot GSB signals become visible immediately after the internal conversion.

The ESA signal is relevant to the optically-allowed transition from the $S_1$ state to the higher excited state. Here only the low-lying $S_2$ and $S_3$ states are considered, whose transition properties along the reaction pathway are given in Table S1-S4. Since these results are rather qualitative and strongly dependent on the electronic structure levels, we do not give the detailed analysis here. But we need to point out that the ESA vanish can be regarded as a signature of the internal conversion.

It is crucial that the results of the population dynamic simulations and the TA PP spectral simulations, taken together, give the logically consistent and fairly complete information on the photoinduced processes in AZM at any level of the electronic structure theory. The following qualitative picture emerges which can be dissected into three steps. (i) The $S_1$ state is populated from the ground in the FC region by the pump pulse and the excited-state wavepacket moves downhill towards lower energies, away from the FC region of the $S_1$ potential energy surface to the area with low values of the $S_0$-$S_1$ TDMs. This step is manifested through the SE signal which quenches on the timescale of several tens of femtoseconds (depending on a specific electronic structure method) when the wavepacket enters (almost) optically dark area of the $S_1$ state. It is important that the SE signal does not correspond to the $S_1$ population dynamics, because the wavepacket remains in the $S_1$ potential energy surface. However, the vanishing of the SE signal directly reflects the time scale at which the wavepacket moves from the FC region. (ii) The arrival of the $S_1$ wavepacket to the low-energy region of the $S_1$ potential energy surface triggers the $S_1 \rightarrow S_0$ internal conversion which is manifested as a hot (negative) GSB wavepacket which exhibits a high-amplitude oscillations in the time domain. (iii) The region of the $S_1$ potential energy surface with weak $S_0$-$S_1$ TDMs possesses strong TDMs from the $S_1$ to higher-lying electronic states. Thus, the ESA can be considered as a reporter of the depopulation of the $S_1$ state and the ESA quenching times roughly coincide with the lifetimes of electronic populations in Figure 1. The onset of the ESA, on the other hand, depends significantly on the electronic structure method employed.

In conclusion, we report the on-the-fly non-adiabatic dynamics simulation of electronic populations and time-resolved TA PP spectra for azomethane. Three different electronic-structure theories are used in the TSH nonadiabatic dynamics simulations. Both OM2/MRCI and SA-CASSCF methods give similar lifetimes of the $S_1$ excited state, while much longer lifetime is obtained at the XMS-CASPT2 level. This is consistent the energy profile of the excited-state reaction pathway from the FC region to the conical intersection. In the simulated TA PP spectra, the SA-CASSCF and XMS-CASPT2 methods give very similar SE signals, and the SE quenching is caused by the decreasing of the $S_0$-$S_1$ TDM along the reaction coordinate, i.e. the CNNC torsion. The OM2/MRCI SE signal is quite different due to the qualitatively different dependence of the TDM on the CNNC torsional angle. The ESA can be regarded as a signature of the internal conversion, since the ESA quenching indicates the completion of the $S_1 \rightarrow S_0$ internal conversion. However, the simulated ESA signals are strongly dependent on the chosen excited-state methods, and this represents a great challenging topic in the electronic-structure theory. Overall, it is demonstrated that electronic populations and time-resolved TA PP signals may reflect different information on the photoinduced excited-state processes. We uncover microscopic reasons for that and establish connections between nonadiabatic dynamics and time-resolved spectroscopic signals. These connections create a bridge between experimental observations and theoretical simulations and deepen our understanding of formation of time-resolved spectra.

   The present work has two important messages. First, the employment of different electronic-structure theories can deeply influence not only the electronic population evolutions but also time-resolved TA PP spectra. Furthermore, different electronic-structure methods can have different impact on these two classes of observables, because evolutions of excited-states populations and SE signals reveal different aspects of the photophysical processes: the population dynamics is governed by the potential energy surface and its gradient, while the TA PP spectrum depends, additionally, on the TDMs. On-the-fly simulations of TA PP signals thus provide an additional and crucial way to examine the reliability of the available electronic-structure methods in the treatment of excited molecular states. Second, TA PP signals may strongly depend on time-dependent changes of TDMs caused by non-Condon effects. Therefore, the cohesive combination of *ab initio* simulations of nonadiabatic dynamics and ultrafast spectroscopic signals can provide a powerful tool to predict and interpret experimental observables, as well as reveal the photoinduced reactions and photophysical mechanisms behind the ultrafast excited-state processes.

## ASSOCIATED CONTENT

## Author Information

### Corresponding Author

E-mail: zhenggang.lan@m.scnu.edu.cn; zhenggang.lan@gmail.com.

### Notes

The authors declare no competing financial interest.

## Acknowledgments

This work is supported by NSFC projects (No. 21933011, 21873112, 21673266 and 21903030). The authors thank the Supercomputing Center, Computer Network Information Center, Chinese Academy of Sciences. M. F. G. acknowledges support from Hangzhou Dianzi University through startup funding.

## Supporting Information

The Supporting Information is available free of charge on the ACS Publications website. Theoretical methods, $S_0$ minimum-energy molecular geometry, transition dipole moment at $S_0$ minimum, transition energies and transition dipole moment along the linear interpolated pathway from $S_0\_min$ to the conical intersection at three methods, GSB, SE, and ESA contributions within 50 fs, time-dependent CNNC dihedral angle distribution for all trajectories in first 100 fs dynamics with different dynamics methods, potential energy curves and the change of transition dipole moments as functions of the NN distance, and the dispersed GSB, SE, and ESA contributions and total signals, and additional implementation details are available.

# Supporting Information

Ultrafast Internal Conversion Dynamics Through the on-the-fly Simulation of Transient Absorption Pump-Probe Spectra with Different Electronic Structure Methods

Chao Xu[1], Kunni Lin[1], Deping Hu[2], Feng Long Gu[1], Maxim F. Gelin[3] and Zhenggang Lan[2,*]

[1] Key Laboratory of Theoretical Chemistry of Environment, Ministry of Education; School of Chemistry, South China Normal University, Guangzhou 510006, P. R. China.

[2] Guangdong Provincial Key Laboratory of Chemical Pollution and Environmental Safety and MOE Key Laboratory of Environmental Theoretical Chemistry, SCNU Environmental Research Institute, School of Environment, South China Normal University, Guangzhou 510006, P. R. China.

[3] School of Sciences, Hangzhou Dianzi University, Hangzhou 310018, P. R. China.

E-mail: zhenggang.lan@m.scnu.edu.cn; zhenggang.lan@gmail.com.

## Theory and Methods

**1. Third-order nonlinear spectral**

In general, theoretical description of femtosecond spectroscopic signals comes from the total Hamiltonian[1]

$$\hat{H}(t) = \hat{H}_M + \hat{H}_F(t) \tag{1}$$

where $\hat{H}_M$ is the molecular Hamiltonian, and

$$\hat{H}_F(t) = -\hat{\boldsymbol{\mu}} \cdot \boldsymbol{E}(t) \tag{2}$$

describes how the system interacts with the external fields in the dipole approximation. Here $\hat{\boldsymbol{\mu}}$ is the transition dipole moment operator, and $\boldsymbol{E}(t)$ is the total electric field of the laser pulses involved. The time-dependent response of the molecular system to the external pulses is fully determined by the third-order polarization[2],

$$\boldsymbol{P}^{(3)}(t) = (i)^3 \int_0^\infty dt_3 \int_0^\infty dt_2 \int_0^\infty dt_1 \boldsymbol{E}(t-t_3)\boldsymbol{E}(t-t_3-t_2)\boldsymbol{E}(t-t_3-t_2-t_1) \times S(t_3,t_2,t_1). \tag{3}$$

Here, the total third-order response function is determined as

$$S(t_3,t_2,t_1) = Tr\{\hat{\boldsymbol{\mu}}^I(t_1+t_2+t_3)[\hat{\boldsymbol{\mu}}^I(t_1+t_2),[\hat{\boldsymbol{\mu}}^I(t_1),[\hat{\boldsymbol{\mu}}^I(0),\hat{\rho}(-\infty)]]]\} \tag{4}$$

$$\hat{\boldsymbol{\mu}}^I(t) = e^{i\hat{H}_M(t)} \hat{\boldsymbol{\mu}} e^{-i\hat{H}_M(t)}$$

where $\hat{\rho}(-\infty)$ is the density operator of the molecule system before the arrival of the laser pulses, at $t = -\infty$.

## 2. Transient-absorption pump-probe spectroscopy

In TA PP spectroscopy, the molecular system is initiated by the pump (*pu*) and interrogated by the delayed probe (*pr*) pulses:

$$E(t) = E_{pu}(t) + E_{pr}(t-\tau)$$
$$E_{pu}(t) = \varepsilon_{pu} A_{pu} E_{pu}(t) e^{ik_{pu}x} e^{-i\omega_{pu}t} + c.c. \qquad (5)$$
$$E_{pr}(t-\tau) = \varepsilon_{pr} A_{pr} E_{pr}(t-\tau) e^{ik_{pr}x} e^{-i\omega_{pr}t} + c.c.$$

Here $\tau$ is the time delay between the pump and probe pulse, $\varepsilon_{pu}$ and $\varepsilon_{pr}$ are the unit vectors of the polarization, $A_{pu}$ and $A_{pr}$ represent amplitudes, $E_{pu}(t)$, $E_{pr}(t)$, $k_{pu}$, $k_{pr}$, $\omega_{pu}$ and $\omega_{pr}$ correspond to envelop functions, wave vectors, and the carrier frequencies, respectively.

The TA PP signal is determined by the third-order polarization $P^{(3)}_{k_{pr}}(\tau,t)$ at the direction $k_{pr}$ or its Fourier transform $P^{(3)}_{k_{pr}}(\tau,\omega)$. Specifically, the integral pump-probe signal is defined as[2],

$$I_{int}(\tau,\omega_{pr}) = \omega_{pr} \operatorname{Im}\{\int_{-\infty}^{\infty} dt E_{pr}(t) e^{i\omega_{pr}t} P^{(3)}_{k_{pr}}(\tau,t)\} \qquad (6)$$

while the dispersed pump-probe signal is determined by the expression[2]

$$I_{dis}(\tau,\omega) = \omega_{pr} \operatorname{Im}\{\varepsilon_{pr}(\omega) P^{(3)}_{k_{pr}}(\tau,\omega)\} \qquad (7)$$

($\varepsilon_{pr}(\omega)$ is the Fourier transform of $E_{pr}(t)$).

## 3. Molecular Hamiltonian

To obtain $I_{int}(\tau,\omega_{pr})$ and $I_{dis}(\tau,\omega)$, the molecular Hamiltonian can be presented in the block-diagonal form:

$$\hat{H}_M = \begin{pmatrix} \hat{H}_0 & \hat{H}_{0,I} & 0 \\ \hat{H}_{I,0} & \hat{H}_I & 0 \\ 0 & 0 & \hat{H}_{II} \end{pmatrix} \qquad (8)$$

Here, $\hat{H}_0$ is the nuclear Hamiltonian of the electronic ground state 0, the Hamiltonian $\hat{H}_I$ describes vibronic dynamics in the manifold of the lower-lying electronic states $\{I\}^{2-5}$ which are in resonance with frequencies of the pump and probe pulse from the ground state, the Hamiltonian $\hat{H}_{II}$ describes vibronic dynamics in the manifold of the higher-lying states $\{II\}$ which have electronic energies around $2\omega_{pu}$ and $2\omega_{pr}$ and can be interrogated by the probe pulse from the manifold $\{I\}$, and, finally, $\hat{H}_{0,I}$ and $\hat{H}_{I,0}$ describe the nondiabatic couplings between the ground state 0 and the lower-lying states $\{I\}$. We can define the corresponding transition dipole moment operators as a sum of the lowering and rising and operators,

$$\hat{\boldsymbol{\mu}} = \hat{\boldsymbol{\mu}}^\uparrow + \hat{\boldsymbol{\mu}}^\downarrow \qquad (9)$$

where

$$\hat{\boldsymbol{\mu}}^\uparrow = \begin{pmatrix} 0 & \boldsymbol{\mu}_{0,I} & 0 \\ 0 & 0 & \boldsymbol{\mu}_{I,II} \\ 0 & 0 & 0 \end{pmatrix}, \quad \hat{\boldsymbol{\mu}}^\downarrow = \begin{pmatrix} 0 & 0 & 0 \\ \boldsymbol{\mu}_{I,0} & 0 & 0 \\ 0 & \boldsymbol{\mu}_{II,I} & 0 \end{pmatrix} \qquad (10)$$

Here $\mu_{0,I}$ is responsible for the $\{0\} \to \{I\}$ transitions and $\mu_{I,II}$ is responsible for the $\{I\} \to \{II\}$ transitions.

Employing the rotating wave approximation (RWA), the system-field interaction Hamiltonian can be rewritten as

$$\hat{H}_F(t) = -\hat{\boldsymbol{\mu}}^\uparrow \boldsymbol{E}_{RWA}(t) - \hat{\boldsymbol{\mu}}^\downarrow \boldsymbol{E}^*_{RWA}(t) \qquad (11)$$

where

$$\boldsymbol{E}_{RWA}(t) = \boldsymbol{\varepsilon}_{pu}A_{pu}E_{pu}(t)e^{-i\boldsymbol{k}_{pu}\boldsymbol{x}}e^{i\omega_{pu}t} + \boldsymbol{\varepsilon}_{pr}A_{pr}E_{pr}(t-\tau)e^{-i\boldsymbol{k}_{pr}\boldsymbol{x}}e^{i\omega_{pr}t} \qquad (12)$$

Finally,

$$\hat{\rho}(-\infty) = \hat{\rho}_{0,0} = |0\rangle\langle 0| \otimes |\nu_0\rangle\langle \nu_0| \qquad (13)$$

where $|0\rangle$ is the electronic ground state and $|\nu_0\rangle$ is the nuclear ground state (the lowest vibrational level).

## 4. Doorway-Window representation

To obtain operational expressions for the simulation of PP signals, a series of approximations are introduced. Here, we focus on the integral signal $I_{int}(\tau,\omega_{pr})$. The dispersed signals can be treated similarly[1], and we give the final formulas $I_{dis}(\tau,\omega)$ at the end of this section.

The doorway-window (DW) approximation is introduced in previous works.[3-6] If the pump and probe pulses are well separated, and $\tau$ is longer than durations of the pump and probe pulses, this approximation gives an exact expression for $I_{int}(\tau,\omega_{pr})$. Second, it is assumed that the laser pulses are short on the nuclear dynamics time scale.[2]

The two approximations are well satisfied for ~ 10 fs pulses, and the integral pump-probe signal can be expressed then as[1]

$$I_{int}(\tau,\omega_{pr}) = \omega_{pr}Tr[\hat{D}_0(\omega_{pu})e^{i\hat{H}_0\tau}\hat{W}_0(\omega_{pr})e^{-i\hat{H}_0\tau} + \hat{D}_I(\omega_{pu})e^{i\hat{H}_1\tau}(\hat{W}_I(\omega_{pr}) - \hat{W}_{II}(\omega_{pr}))e^{-i\hat{H}_1\tau}] \qquad (14)$$

where

$$\begin{aligned}\hat{D}_0(\omega_{pu}) &= \int_{-\infty}^{\infty} dt_2' \int_0^{\infty} dt_1 E_{pu}(t_2')E_{pu}(t_2'-t_1)e^{i\omega_{pu}t_1}e^{-i\hat{H}_1 t_1}\boldsymbol{\varepsilon}_{pu}\boldsymbol{\mu}_{1,0}\hat{\rho}_{0,0}e^{i\hat{H}_0 t_1}\boldsymbol{\varepsilon}_{pu}\boldsymbol{\mu}_{0,1} + H.c. \\ \hat{D}_I(\omega_{pu}) &= \int_{-\infty}^{\infty} dt_2' \int_0^{\infty} dt_1 E_{pu}(t_2')E_{pu}(t_2'-t_1)e^{i\omega_{pu}t_1}\boldsymbol{\varepsilon}_{pu}\boldsymbol{\mu}_{0,1}e^{-i\hat{H}_1 t_1}\boldsymbol{\varepsilon}_{pu}\boldsymbol{\mu}_{1,0}\hat{\rho}_{0,0}e^{i\hat{H}_0 t_1} + H.c.\end{aligned} \qquad (15)$$

are the D-operators and

$$\hat{W}_0(\omega_{pr}) = \int_{-\infty}^{\infty} dt' \int_0^{\infty} dt_3 E_{pr}(t') E_{pr}(t'+t_3) e^{i\omega_{pr}t_3} e^{i\hat{H}_0 t_3} \boldsymbol{\varepsilon}_{pr}\boldsymbol{\mu}_{0,\mathrm{I}} e^{-i\hat{H}_\mathrm{I} t_3} \boldsymbol{\varepsilon}_{pr}\boldsymbol{\mu}_{\mathrm{I},0} + H.c.$$

$$\hat{W}_\mathrm{I}(\omega_{pr}) = \int_{-\infty}^{\infty} dt' \int_0^{\infty} dt_3 E_{pr}(t') E_{pr}(t'+t_3) e^{i\omega_{pr}t_3} \boldsymbol{\varepsilon}_{pr}\boldsymbol{\mu}_{\mathrm{I},0} e^{i\hat{H}_0 t_3} \boldsymbol{\varepsilon}_{pr}\boldsymbol{\mu}_{0,\mathrm{I}} e^{-i\hat{H}_\mathrm{I} t_3} + H.c. \quad (16)$$

$$\hat{W}_\mathrm{II}(\omega_{pr}) = \int_{-\infty}^{\infty} dt' \int_0^{\infty} dt_3 E_{pr}(t') E_{pr}(t'+t_3) e^{i\omega_{pr}t_3} \boldsymbol{\varepsilon}_{pr}\boldsymbol{\mu}_{\mathrm{I},\mathrm{II}} e^{-i\hat{H}_\mathrm{II} t_3} \boldsymbol{\varepsilon}_{pr}\boldsymbol{\mu}_{\mathrm{II},\mathrm{I}} e^{i\hat{H}_\mathrm{I} t_3} + H.c.$$

are the W-operators. As defined in Eq. (14), the terms $\hat{D}_0 \hat{W}_0$, $\hat{D}_\mathrm{I} \hat{W}_\mathrm{I}$, $\hat{D}_\mathrm{I} \hat{W}_\mathrm{II}$ are proportional to the ground-state bleach (GSB), stimulated emission (SE), and excited-state absorption (ESA) contributions to the signal, correspondingly.

## 5. Quasi-classical evaluation

$I_{\mathrm{int}}(\tau,\omega_{pr})$ in Eq. (14) is fully quantum. To evaluate it by classical trajectories, we resort to the following approximations. In the phase space, the DW operators are treated as functions of nuclear coordinates. The trace is replaced by the integral over the nuclear phase space. Finally, the quantum evolutions are approximated by evolutions along classical trajectories. Thus,

$$\begin{aligned}
I_{\mathrm{int}}(\tau,\omega_{pr}) &= I_{\mathrm{int}}^{GSB}(\tau,\omega_{pr}) + I_{\mathrm{int}}^{SE}(\tau,\omega_{pr}) + I_{\mathrm{int}}^{ESA}(\tau,\omega_{pr}) \\
I_{\mathrm{int}}^{GSB}(\tau,\omega_{pr}) &= \omega_{pr} \int d\boldsymbol{R}_g d\boldsymbol{P}_g D_0(\omega_{pu},\boldsymbol{R}_g,\boldsymbol{P}_g) W_0^{\mathrm{int}}(\omega_{pr},\boldsymbol{R}_g(\tau),\boldsymbol{P}_g(\tau)) \\
I_{\mathrm{int}}^{SE}(\tau,\omega_{pr}) &= \omega_{pr} \int d\boldsymbol{R}_g d\boldsymbol{P}_g D_\mathrm{I}(\omega_{pu},\boldsymbol{R}_g,\boldsymbol{P}_g) W_\mathrm{I}^{\mathrm{int}}(\omega_{pr},\boldsymbol{R}_e(\tau),\boldsymbol{P}_e(\tau)) \\
I_{\mathrm{int}}^{ESA}(\tau,\omega_{pr}) &= -\omega_{pr} \int d\boldsymbol{R}_g d\boldsymbol{P}_g D_\mathrm{I}(\omega_{pu},\boldsymbol{R}_g,\boldsymbol{P}_g) W_\mathrm{II}^{\mathrm{int}}(\omega_{pr},\boldsymbol{R}_e(\tau),\boldsymbol{P}_e(\tau))
\end{aligned} \quad (17)$$

Here, $\boldsymbol{R}_g$ and $\boldsymbol{P}_g$ are the initial nuclear coordinates and momenta in the electronic ground state, which are sampled from the Wigner distribution $\rho_g^{Wig}(\boldsymbol{R}_g,\boldsymbol{P}_g)$, $\boldsymbol{R}_g(\tau)$ and $\boldsymbol{P}_g(\tau)$ are the coordinates and momenta after evolution in the ground state up to $t=\tau$, and $\boldsymbol{R}_e(\tau)$ and $\boldsymbol{P}_e(\tau)$ are the coordinates and momenta after propagation in the manifold of lower-lying excited states {I} up to $t=\tau$.

Finally, the semiclassical doorway functions are defined as

$$D_0(\omega_{pu}, \mathbf{R}_g, \mathbf{P}_g) = \sum_e |\boldsymbol{\varepsilon}_{pu} \boldsymbol{\mu}_{ge}(\mathbf{R}_g)|^2 E_{pu}^2[\omega_{pu} - U_{eg}(\mathbf{R}_g)] \rho_g^{Wig}(\mathbf{R}_g, \mathbf{P}_g)$$
$$D_I(\omega_{pu}, \mathbf{R}_g, \mathbf{P}_g) = |\boldsymbol{\varepsilon}_{pu} \boldsymbol{\mu}_{ge}(\mathbf{R}_g)|^2 E_{pu}^2[\omega_{pu} - U_{eg}(\mathbf{R}_g)] \rho_g^{Wig}(\mathbf{R}_g, \mathbf{P}_g)$$
(18)

and the semiclassical W-functions are expressed as

$$W_0^{int}(\omega_{pr}, \mathbf{R}_g(\tau), \mathbf{P}_g(\tau)) = \sum_e |\boldsymbol{\varepsilon}_{pr} \boldsymbol{\mu}_{ge}(\mathbf{R}_g(\tau))|^2 E_{pr}^2[\omega_{pr} - U_{eg}(\mathbf{R}_g(\tau))]$$
$$W_I^{int}(\omega_{pr}, \mathbf{R}_e(\tau), \mathbf{P}_e(\tau)) = |\boldsymbol{\varepsilon}_{pr} \boldsymbol{\mu}_{ge(\tau)}(\mathbf{R}_e(\tau))|^2 E_{pr}^2[\omega_{pr} - U_{e(\tau)g}(\mathbf{R}_e(\tau))]$$
(19)
$$W_{II}^{int}(\omega_{pr}, \mathbf{R}_e(\tau), \mathbf{P}_e(\tau)) = \sum_f |\boldsymbol{\varepsilon}_{pr} \boldsymbol{\mu}_{e(\tau)f}(\mathbf{R}_e(\tau))|^2 E_{pr}^2[\omega_{pr} - U_{fe(\tau)}(\mathbf{R}_e(\tau))]$$

In the above formulas, $\varepsilon_{pu}(\omega_{pu})$ and $\varepsilon_{pr}(\omega_{pr})$ are the Fourier transforms of the $E_{pu}(t)$ and $E_{pr}(t)$; the ground electronic state is denoted as $g$; electronic states of the manifold {I} are marked as $e$; electronic states of the manifold {II} are denoted as $f$; $\boldsymbol{\mu}_{ge}(\mathbf{R}_g(\tau))$, $\boldsymbol{\mu}_{ge(\tau)}(\mathbf{R}_e(\tau))$ and $\boldsymbol{\mu}_{e(\tau)f}(\mathbf{R}_e(\tau))$ are the transition dipole moments between the corresponding electronic states for a specific nuclear configuration $\mathbf{R}_g(\tau)$ or $\mathbf{R}_e(\tau)$; $U_{eg}(\mathbf{R}_e(\tau))$, $U_{e(\tau)g}(\mathbf{R}_e(\tau))$ and $U_{fe(\tau)}(\mathbf{R}_e(\tau))$ are the energy gaps between the corresponding electronic states for a specific nuclear configuration $\mathbf{R}_g(\tau)$ or $\mathbf{R}_e(\tau)$. The notion $e(\tau)$ means that a trajectory initiated at $t = 0$ in a state $e$ of the manifold {I} can end up at time $\tau$ in another state of the manifold {I} or in the ground state {0}.

If we allow for the $e(\tau) \to g$ internal conversion (IC), that is allow the trajectory to jump from $e(\tau)$ to $g$, the signal is defined as

$$S_{int}(\tau,\omega_{pr})=S_{int}^{GSB}(\tau,\omega_{pr})+S_{int}^{SE}(\tau,\omega_{pr})+S_{int}^{ESA}(\tau,\omega_{pr})$$

$$S_{int}^{GSB}(\tau,\omega_{pr}) = \int d\mathbf{R}_g d\mathbf{P}_g \hat{D}_{0,IC}(\omega_{pu},\mathbf{R}_g,\mathbf{P}_g)\hat{W}_{0,IC}^{int}(\omega_{pr},\omega,\mathbf{R}_g(\tau),\mathbf{P}_g(\tau))$$

$$S_{int}^{SE}(\tau,\omega_{pr}) = \int d\mathbf{R}_g d\mathbf{P}_g \hat{D}_{I,IC}(\omega_{pu},\mathbf{R}_e,\mathbf{P}_e)\hat{W}_{I,IC}^{int}(\omega_{pr},\omega,\mathbf{R}_e(\tau),\mathbf{P}_e(\tau))$$

$$S_{int}^{ESA}(\tau,\omega_{pr}) = -\int d\mathbf{R}_g d\mathbf{P}_g \hat{D}_{I,IC}(\omega_{pu},\mathbf{R}_e,\mathbf{P}_e)\hat{W}_{II,IC}^{int}(\omega_{pr},\omega,\mathbf{R}_e(\tau),\mathbf{P}_e(\tau))$$

(20)

Here

$$D_{0,IC}(\omega_{pu},\mathbf{R}_g,\mathbf{P}_g) = \begin{cases} D_0(\omega_{pu},\mathbf{R}_g,\mathbf{P}_g), \text{if trajectory stays within \{0\}} \\ 0, \text{if trajectory stays within \{I\}} \\ -D_I(\omega_{pu},\mathbf{R}_g,\mathbf{P}_g), \text{if trajectory jumps from \{I\} to \{0\}} \end{cases}$$

(21)

$$D_{I,IC}(\omega_{pu},\mathbf{R}_e,\mathbf{P}_e) = \begin{cases} 0, \text{if trajectory stays within \{0\}} \\ D_I(\omega_{pu},\mathbf{R}_e,\mathbf{P}_e), \text{if trajectory stays within \{I\}} \\ 0, \text{if trajectory jumps from \{I\} to \{0\}} \end{cases}$$

(22)

and

$$W_{0,IC}^{int}(\omega_{pr},\omega,\mathbf{R}_g(\tau),\mathbf{P}_g(\tau)) = \begin{cases} W_0^{int}(\omega_{pr},\omega,\mathbf{R}_g(\tau),\mathbf{P}_g(\tau)), \text{ if trajectory stays within \{0\}} \\ 0, \text{ if trajectory stays within \{I\}} \\ W_0^{int}(\omega_{pr},\omega,\mathbf{R}_g(\tau),\mathbf{P}_g(\tau)), \text{ if trajectory jumps from \{I\} to \{0\}} \end{cases}$$

(23)

$$W_{I,IC}^{int}(\omega_{pr},\omega,\mathbf{R}_e(\tau),\mathbf{P}_e(\tau)) = \begin{cases} 0, \text{ if trajectory stays within \{0\}} \\ W_I^{int}(\omega_{pr},\omega,\mathbf{R}_e(\tau),\mathbf{P}_e(\tau)), \text{ if trajectory stays within \{I\}} \\ 0, \text{ if trajectory jumps from \{I\} to \{0\}} \end{cases}$$

(24)

$$W_{II,IC}^{int}(\omega_{pr},\omega,\mathbf{R}_e(\tau),\mathbf{P}_e(\tau)) = \begin{cases} 0, \text{ if trajectory stays within \{0\}} \\ W_{II}^{int}(\omega_{pr},\omega,\mathbf{R}_e(\tau),\mathbf{P}_e(\tau)), \text{ if trajectory stays within \{I\}} \\ 0, \text{ if trajectory jumps from \{I\} to \{0\}} \end{cases}$$

(25)

The semiclassical formulas for the dispersed pump-probe signal read as follows[1]

$$I_{dis}(\tau,\omega) = I_{dis}^{GSB}(\tau,\omega_{pr}) + I_{dis}^{SE}(\tau,\omega_{pr}) + I_{dis}^{ESA}(\tau,\omega_{pr})$$

$$I_{dis}^{GSB}(\tau,\omega_{pr}) = \omega_{pr}\int d\mathbf{R}_g d\mathbf{P}_g D_0(\omega_{pu},\mathbf{R}_g,\mathbf{P}_g) W_0^{dis}(\omega_{pr},\omega,\mathbf{R}_g(\tau),\mathbf{P}_g(\tau))$$

$$I_{dis}^{SE}(\tau,\omega_{pr}) = \omega_{pr}\int d\mathbf{R}_g d\mathbf{P}_g D_I(\omega_{pu},\mathbf{R}_g,\mathbf{P}_g) W_I^{dis}(\omega_{pr},\omega,\mathbf{R}_e(\tau),\mathbf{P}_e(\tau))$$

$$I_{dis}^{ESA}(\tau,\omega_{pr}) = -\omega_{pr}\int d\mathbf{R}_g d\mathbf{P}_g D_I(\omega_{pu},\mathbf{R}_g,\mathbf{P}_g) W_{II}^{dis}(\omega_{pr},\omega,\mathbf{R}_e(\tau),\mathbf{P}_e(\tau))$$

(26)

Here the D-functions are given by Eqs. (18) and (19), while the W-functions are defined as

$$W_0^{dis}(\omega_{pr},\omega,\mathbf{R}_g(\tau),\mathbf{P}_g(\tau)) = \varepsilon_{pr}^2[\omega-\omega_{pr}]\sum_e |\boldsymbol{\mu}_{ge}(\mathbf{R}_g(\tau))|^2 \frac{\upsilon}{\upsilon^2+[\omega-U_{eg}(\mathbf{R}_g(\tau))]^2}$$

$$W_I^{dis}(\omega_{pr},\omega,\mathbf{R}_e(\tau),\mathbf{P}_e(\tau)) = \varepsilon_{pr}^2[\omega-\omega_{pr}]|\boldsymbol{\mu}_{ge(\tau)}(\mathbf{R}_e(\tau))|^2 \frac{\upsilon}{\upsilon^2+[\omega-U_{e(\tau)g}(\mathbf{R}_e(\tau))]^2}$$

$$W_{II}^{dis}(\omega_{pr},\omega,\mathbf{R}_e(\tau),\mathbf{P}_e(\tau)) = \varepsilon_{pr}^2[\omega-\omega_{pr}]\sum_f |\boldsymbol{\mu}_{e(\tau)f}(\mathbf{R}_e(\tau))|^2 \frac{\upsilon}{\upsilon^2+[\omega-U_{fe(\tau)}(\mathbf{R}_e(\tau))]^2}$$

(27)

where $\upsilon$ is the electronic dephasing rate.

Once the dispersed signal is detected by a spectrometer, which means it is not heterodyned with the probe pulse, we obtain $I_{dis}(\tau,\omega) \sim P(\tau,\omega)$. In this case, one has to replace $E_{pr}^2(\omega-\omega_{pr})$ with $E_{pr}(\omega-\omega_{pr})$ in Eq. (27). The impulsive dispersed PP signal, which has perfect time and frequency resolution,[7] is obtained by setting $E_{pr}^2(\omega-\omega_{pr}) \to 1$ in Eq. (27).

## 3. On-the-fly non-adiabatic dynamics

In the current work, the non-adiabatic dynamics of azomethane was simulated by the on-the-fly trajectory surface hopping (TSH) approach[8] at different electronic-structure levels, including OM2/MRCI[9,10], SA-CASSCF[11,12], and XMS-CASPT2[13-15]. The ground state minimum ($S_0$) was optimized at the B3LYP/6-31G*[16] level by using Gaussian 16.[17] The conical intersection (two states were included) was optimized at OM2/MRCI, SA-CASSCF and XMS-CASPT2 level, respectively. The initial conditions of the nuclear coordinates and momenta were given by Wigner sampling[18] of the lowest vibrational level on the electronic ground state. To directly compare the results of dynamics simulations and TA PP signals, the same initial samplings (nuclear coordinates and momenta) were used for all dynamics calculations. All trajectories were propagated up to 400 fs. The time steps in the propagation of the nuclear and electronic motion were 0.5 fs and 0.005 fs, respectively.

All details of these electronic-structure calculations are explained below.

(1) OM2/MRCI: At the semiempirical OM2/MRCI level, the active space includes 8 active electrons in 7 orbitals. The OM2 Hamiltonian used the ROHF (restricted open-shell Hartree-Fock) in the construction of the reference orbitals. In the MRCI calculations, all electronic configurations in the GUGA-CI calculation were generated from three reference configurations, namely closed-shell, single, and double HOMO-LUMO excitations. All OM2/MRCI calculations were performed by the MNDO package.[19] The interface between the non-adiabatic dynamics in the JADE package[20] and OM2/MRCI calculations in the MNDO package was employed in the on-the-fly TSH dynamics. The non-adiabatic dynamics propagation included the two lowest electronic states. After it, the ESA spectral calculations involve the higher-excited states, and four states (including ground state and three excited states) were considered here.

(2) SA-CASSCF: In the SA-CASSCF calculation, the active space CAS (6,4) consisted of $n$, $\pi$, $\pi^*$ orbitals and three-state averaged calculations were performed. The CASSCF electronic structure calculations by the Molpro 2020 package[21]. The interface between the non-adiabatic dynamics in the JADE package[22] and CASSCF calculations in the Molpro package was employed in the on-the-fly TSH dynamics. To obtain the ESA signals, the four electronic states (including ground state and three excited states) were considered.

(3) XMS-CASPT2: In the XMS-CASPT2 calculations, the same active state used in the SA

(3)-CASSCF calculations were taken. The interface between the non-adiabatic dynamics in the JADE package[23] and XMS-CASPT2 calculations in the BAGEL package[24] was employed in the on-the-fly TSH dynamics. To obtain the ESA signals, the four electronic states (including ground state and three excited states) were calculated.

## Supplementary Data

The $S_0$ minimum-energy molecular geometry of azomethane was optimized at the B3LYP/6-31G* level, as shown in Figure S1. The $r_{NN}$ of the N-N bond distance is 1.243 Å, and the CNNC dihedral angle $\tau_{CNNC}$ is 0.0 degrees. The potential energies and transition dipole moments (TDMs) between different electronic states at $S_0$-min calculated by three electronic-structure methods are given in Table S1 and Table S2. At $S_0$ geometry, the $S_1$ energy is predicted at 3.40 eV (OM2/MRCI), 3.55 eV (SA-CASSCF) and 3.48 eV (XMS-CASPT2). At all levels of theory, the $S_1$ state is an excited state corresponding to the $n$-$\pi$* transition. The excitation energy of the $S_1$ state at the SA-CASSCF level is slightly higher than those at the other two levels, because the SA-CASSCF method does not capture dynamical electronic correlations. Conical intersections located by three different methods display significant torsion of the dihedral angle CNNC. At the conical intersection, the dihedral angle C-N-N-C is around -98.3, -92.8, and -90.5 degrees, and the N-N bond is 1.217, 1.269, and 1.271 Å at the OM2/MRCI, SA-CASSCF, and XMS-CASPT2 levels, respectively.

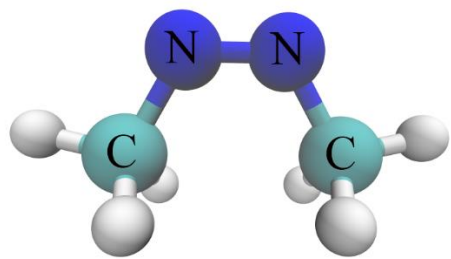

**Figure S1.** Optimized $S_0$ geometry of azomethane with atom labels.

**Table S1**. Transition dipole moments (a.u.) at $S_0$ minimum.

|  | OM2/MRCI | SA-CASSCF | XMS-CASPT2 |
| --- | --- | --- | --- |
| $S_0 \rightarrow S_1$ | 0.04227 | 0.29394 | 0.16890 |
| $S_0 \rightarrow S_2$ | 0.00035 | 0.03900 | 0.00021 |
| $S_0 \rightarrow S_3$ | 0.03657 | 0.04080 | 0.02531 |
| $S_1 \rightarrow S_2$ | 0.33257 | 0.36279 | 0.25137 |
| $S_1 \rightarrow S_3$ | 0.00012 | 0.35134 | 0.21131 |
| $S_2 \rightarrow S_3$ | 0.05854 | 0.00446 | 0.00009 |

**Table S2**. Potential energies (eV) and transition dipole moments (a.u.) along the linear interpolated pathway from $S_0\_min$ to the conical intersection at OM2/MRCI level. The conical intersection was optimized at OM2/MRCI level and two states were included.

| LIIC | $S_0$ | $S_1$ | $S_2$ | $S_3$ | $S_1 \rightarrow S_2$ | $S_1 \rightarrow S_3$ |
|------|-------|-------|-------|-------|-----------------------|-----------------------|
| 0    | 0.00  | 3.38  | 5.78  | 6.76  | 0.33257               | 0.00012               |
| 1    | 0.02  | 3.32  | 5.80  | 6.72  | 0.33928               | 0.02472               |
| 2    | 0.10  | 3.21  | 5.83  | 6.71  | 0.33785               | 0.03604               |
| 3    | 0.24  | 3.07  | 5.84  | 6.38  | 0.30694               | 0.16066               |
| 4    | 0.43  | 2.91  | 5.63  | 6.17  | 0.18740               | 0.27841               |
| 5    | 0.66  | 2.75  | 5.15  | 6.20  | 0.14073               | 0.28366               |
| 6    | 0.94  | 2.60  | 4.61  | 6.30  | 0.14102               | 0.25658               |
| 7    | 1.26  | 2.47  | 4.06  | 6.44  | 0.14868               | 0.22684               |
| 8    | 1.62  | 2.36  | 3.53  | 6.62  | 0.16299               | 0.22073               |
| 9    | 1.98  | 2.29  | 3.06  | 6.81  | 0.18850               | 0.18871               |
| 10   | 2.22  | 2.22  | 2.80  | 7.02  | 0.17503               | 0.07831               |

**Table S3**. Potential energies (eV) and transition dipole moments (a.u.) along the linear interpolated pathway from $S_0$_min to the conical intersection at SA-CASSCF level. The conical intersection was optimized at SA-CASSCF level and two states were included.

| LIIC | $S_0$ | $S_1$ | $S_2$ | $S_3$ | $S_1 \rightarrow S_2$ | $S_1 \rightarrow S_3$ |
|---|---|---|---|---|---|---|
| 0  | 0.00 | 3.62 | 6.21 | 7.52 | 0.34061 | 0.33583 |
| 1  | 0.04 | 3.58 | 6.13 | 7.40 | 0.33523 | 0.33049 |
| 2  | 0.13 | 3.47 | 6.02 | 7.21 | 0.32234 | 0.31882 |
| 3  | 0.30 | 3.36 | 5.85 | 7.00 | 0.29947 | 0.30562 |
| 4  | 0.53 | 3.24 | 5.61 | 6.83 | 0.26719 | 0.29298 |
| 5  | 0.81 | 3.12 | 5.29 | 6.71 | 0.23230 | 0.27838 |
| 6  | 1.12 | 3.00 | 4.92 | 6.65 | 0.19551 | 0.28410 |
| 7  | 1.47 | 2.88 | 4.54 | 6.63 | 0.16347 | 0.29241 |
| 8  | 1.89 | 2.87 | 4.11 | 6.67 | 0.15917 | 0.30325 |
| 9  | 2.23 | 2.74 | 3.85 | 6.70 | 0.18319 | 0.31099 |
| 10 | 2.48 | 2.55 | 3.80 | 6.74 | 0.22314 | 0.28104 |

**Table S4**. Potential energies (eV) and transition dipole moments (a.u.) along the linear interpolated pathway from $S_0\_min$ to the conical intersection at XMS-CASPT2 level. The conical intersection was optimized at XMS-CASPT2 level and two states were included.

| LIIC | $S_0$ | $S_1$ | $S_2$ | $S_3$ | $S_1 \rightarrow S_2$ | $S_1 \rightarrow S_3$ |
|------|------|------|------|------|---------|---------|
| 0    | 0.00 | 3.48 | 6.28 | 7.37 | 0.36279 | 0.35134 |
| 1    | 0.03 | 3.41 | 6.23 | 7.22 | 0.35634 | 0.34875 |
| 2    | 0.14 | 3.32 | 6.14 | 6.99 | 0.33370 | 0.34588 |
| 3    | 0.30 | 3.20 | 5.98 | 6.75 | 0.29938 | 0.34623 |
| 4    | 0.53 | 3.08 | 5.67 | 6.57 | 0.25383 | 0.34819 |
| 5    | 0.82 | 2.96 | 5.31 | 6.50 | 0.20812 | 0.34483 |
| 6    | 1.13 | 2.87 | 4.87 | 6.33 | 0.17428 | 0.34710 |
| 7    | 1.45 | 2.77 | 4.43 | 6.35 | 0.15605 | 0.31955 |
| 8    | 1.83 | 2.70 | 4.00 | 6.41 | 0.15094 | 0.31285 |
| 9    | 2.19 | 2.62 | 3.67 | 6.49 | 0.15312 | 0.30314 |
| 10   | 2.45 | 2.51 | 3.54 | 6.58 | 0.06431 | 0.28319 |

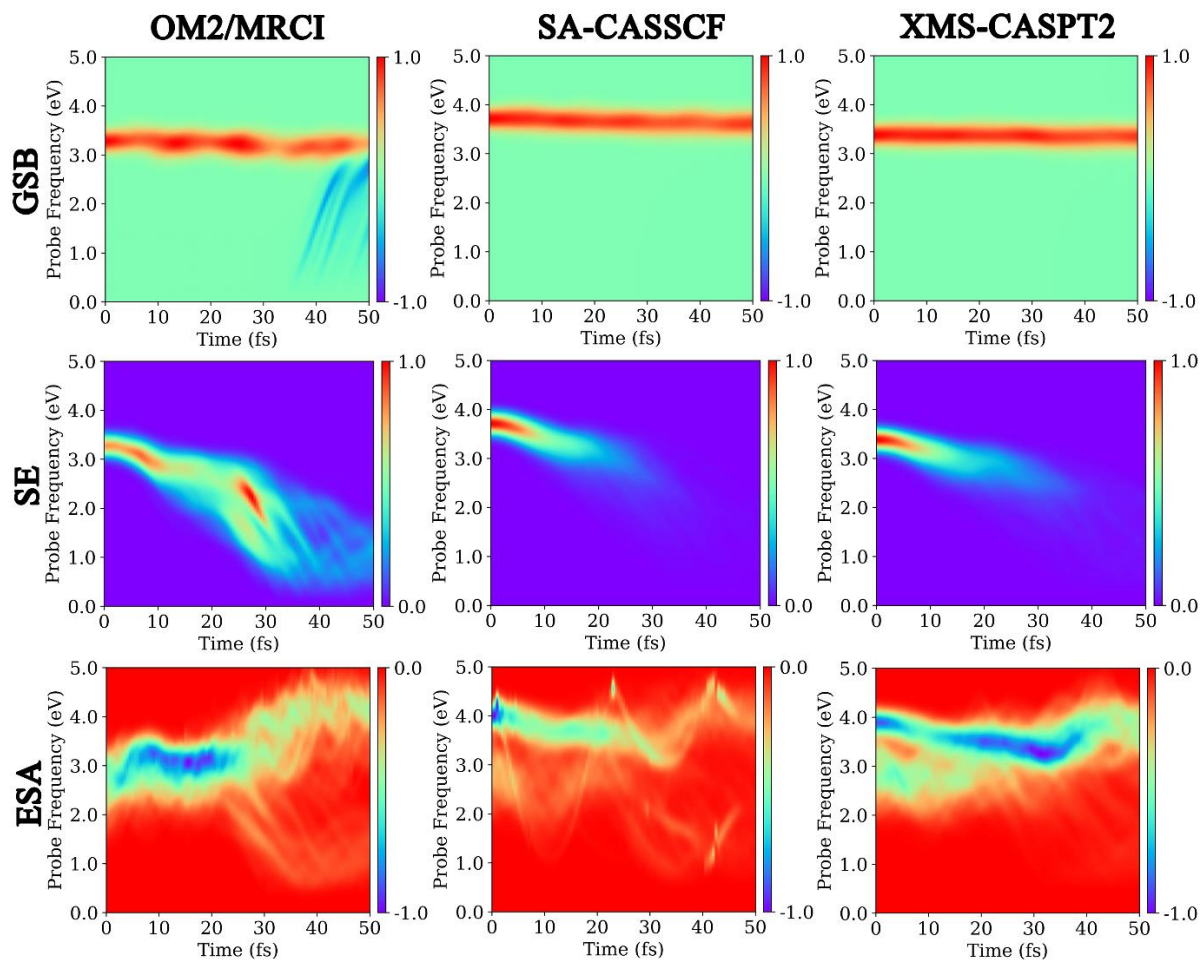

**Figure S2.** Normalized GSB, SE, and ESA contributions and total integral signal as a function of the pump-probe delay time T and the central frequency $\omega_{pr}$ of the probe pulse. The duration of both pump and probe pulses are $\tau_a = 5$ fs. The central frequency $\omega_{pu}$ of the pump pulse is tuned into resonance with the $S_1(n\pi^*)$ state (OM2/MRCI: 3.37 eV, SA-CASSCF: 3.70 eV, XMS-CASPT2: 3.40 eV). Left column: OM2/MRCI method, Middle column: SA-CASSCF method, Right column: XMS-CASPT2 method.

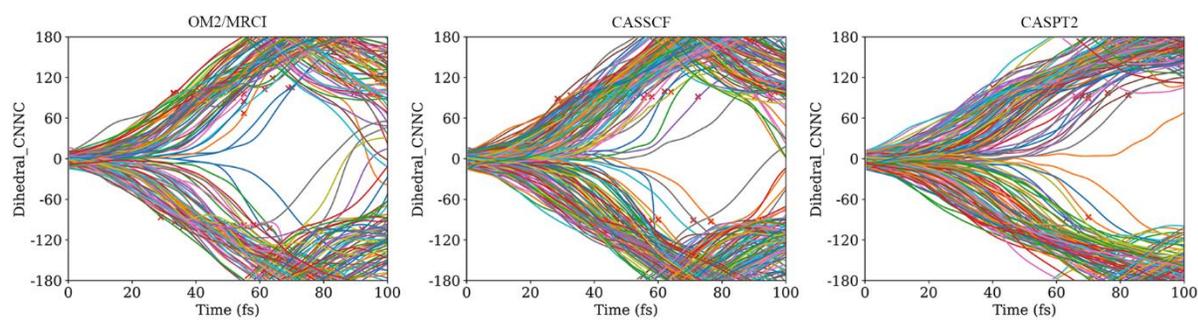

**Figure S3.** Time-dependent CNNC dihedral angle distributions evaluated by three different dynamics methods during first 100 fs.

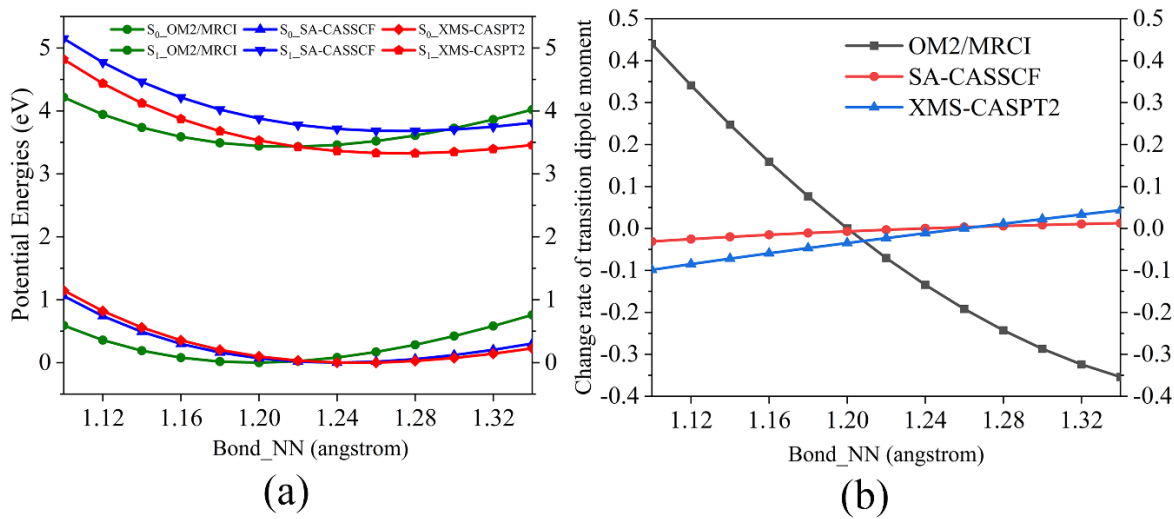

**Figure S4**. Potential energy curves (a) and the change of $S_0 \rightarrow S_1$ transition dipole moments (b) as functions of the NN distance. The change of the transition dipole moment at the optimized geometry is set to zero.

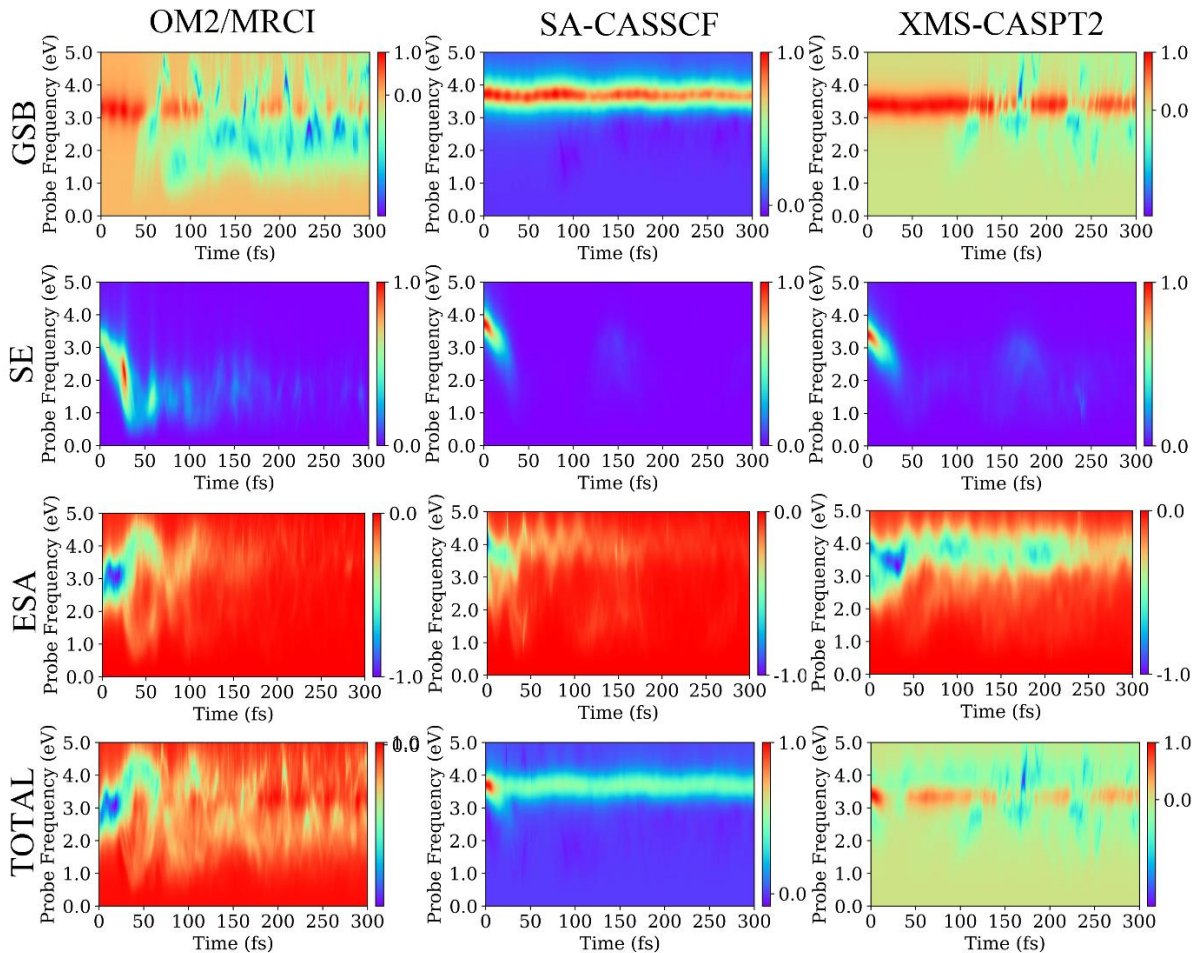

**Figure S5.** Normalized GSB, SE, and ESA contributions and total dispersed signal as a function of the pump- probe delay time T and the central frequency $\omega_{pr}$ of the probe pulse. The durations of both pump and probe pulses are $\tau_a = 5$ fs, and $v = 0.01$ eV. The central frequency $\omega_{pu}$ of the pump pulse is tuned into resonance with the $S_1(n\pi^*)$ state. Left column: OM2/MRCI method, Middle column: SA-CASSCF method, Right column: XMS-CASPT2 method.